\documentclass[twocolumn,english,preprintnumbers,prb]{revtex4-1}
\usepackage[latin9]{inputenc}
\usepackage{amsmath}
\usepackage{amssymb}
\usepackage{graphicx}
\usepackage{esint}
\usepackage{natbib}
\usepackage{hyperref}
\hypersetup{colorlinks=true,linkcolor=blue,filecolor=magenta,urlcolor=blue,citecolor=blue,}

\makeatletter

\@ifundefined{textcolor}{}{%
 \definecolor{BLACK}{gray}{0}
 \definecolor{WHITE}{gray}{1}
 \definecolor{RED}{rgb}{1,0,0}
 \definecolor{GREEN}{rgb}{0,1,0}
 \definecolor{BLUE}{rgb}{0,0,1}
 \definecolor{CYAN}{cmyk}{1,0,0,0}
 \definecolor{MAGENTA}{cmyk}{0,1,0,0}
 \definecolor{YELLOW}{cmyk}{0,0,1,0}
}

\usepackage{color}
\usepackage{ulem}

\usepackage{aecompl}

\usepackage{epsfig}\usepackage{dcolumn}\usepackage{bm}

\usepackage{babel}

\makeatother

\usepackage{babel}
\begin{document}

\title{Resonant electron tunneling spectroscopy of antibonding states in a Dirac semimetal}

\author{Y. Marques$^{1}$, D. Yudin$^{2}$, I. A. Shelykh$^{2,3}$, and
A. C. Seridonio$^{1,4}$}

\affiliation{$^{1}$Departamento de F\'{i}sica e Qu\'{i}mica, Unesp - Univ Estadual Paulista, 15385-000, Ilha Solteira, SP, Brazil\\
$^{2}$ITMO University, St. Petersburg 197101, Russia\\
$^{3}$Science Institute, University of Iceland, Dunhagi-3, IS-107, Reykjavik, Iceland\\
$^{4}$IGCE, Unesp - Univ Estadual Paulista, Departamento de F\'{i}sica, 13506-900, Rio Claro, SP, Brazil}

\begin{abstract}
Recently, it was shown both theoretically and experimentally that certain three-dimensional (3D) materials have Dirac points in the Brillouin zone, thus being 3D analogs of graphene. Moreover, it was suggested that under specific conditions a pair of localized impurities placed inside a three-dimensional Dirac semimetal may lead to the formation of an unusual antibonding state. In the meantime, the effect of vibrational degrees of freedom which are present in any realistic system has avoided attention. In this work, we address the influence of phonons on characteristic features of (anti)bonding state, and discuss how these results can be tested experimentally via local probing, namely, inelastic electron tunneling spectroscopy curve obtained in STM measurements.
\end{abstract}

\maketitle

\section{Introduction}

Quantum electrodynamics (QED) is an abundant source of ideas and concepts that form a basis of the modern science \cite{Fujikawa2004}. However, direct experimental observation of many effects predicted by QED can hardly be accomplished as it requires achievement of the energies which can not be reached in up-to-date experimental setups. Fortunately, in the domain of condensed matter physics there exist low-energy analogs of a plethora of QED phenomena. One of the most striking examples is Weyl fermions originally proposed as massless chiral particles mathematically described by Dirac equation \cite{Weyl1929} and recently shown to emerge in a certain class of semimetals with non-trivial topology \cite{Bevan1997,Jiang2012,Halasz2012,Manes2012,Xu2015,Huang2015,Armitage2018}.

In most of the cases the spectrum of the electrons and the holes in metals and semiconductors can be well approximated by parabolic dispersion relation for both conduction and valence bands (so-called effective mass approximation). In certain class of materials, known as topological insulators, conduction and valence bands become inverted. Transition between topological and trivial phases proceeds via a gapless phase which is characterized by three-dimensional Dirac dispersion in the presence of both time-reversal and inversion symmetry \cite{Murakami2007}. Interestingly, in the systems with a lack of inversion symmetry the presence of such critical point results in the formation of gapless Weyl nodes, distributed over the Brillouin zone and annihilating with partners of opposite chirality, leading thus to the change in topology \cite{Volovik2003} and appearance of the exotic surface states in the form of Fermi arcs \cite{Wan2011}. For the materials with the chemical potential positioned quite close to the Weyl nodes the term Weyl semimetal has been coined, in which electronic structure is characterized by bulk band crossings.

In three-dimensional (3D) materials, dispersion in the form of massless Dirac equation with four-fold degenerate Dirac points was predicted in Refs. \onlinecite{Abrikosov1971,Lee2012,Young2012}. In contrast to Weyl nodes, this degeneracy is not topologically protected owing to the fact that the net Chern number is zero. In certain cases, however, the space group of a crystalline solid prevents this mixing to happen, and symmetry-protected nodes appear. This corresponds to the case of three-dimensional Dirac semimetals (3D-DSM) \cite{Wan2011,Burkov2011,Young2012} existing in the vicinity of the transition point between topological and normal phases and stabilized by crystalline symmetry. In these materials, valence and conduction bands touch at discrete Dirac points, emerging due to the overlap of the Weyl nodes in momentum space and characterized by linear electronic dispersion in all three directions. The latter makes such structures extremely robust with respect to external perturbations, e.g., spin-orbit coupling. The existence of three-dimensional Dirac points has been predicted and experimentally verified in Na$_3$Bi \cite{Wang2012,Liu2014a} and Cd$_3$As$_2$ \cite{Wang2013,Liu2014b}.

Recently, it was shown that for a pair of distant localized impurities embedded inside the 3D-DSM one can expect the formation of antibonding ground state \cite{Marques2017}. In this paper, we generalize our previous study \cite{Marques2017} to explore the effect of vibrational modes on the ground-state formation. We expect that the peculiarities associated with the presence of phonons can be detected by means of inelastic electron tunneling spectroscopy (IETS) with the help of local probing technique, e.g., scanning tunneling microscopy (STM). In fact, when the energy of the tunneling electron exceeds the energy required to excite the local vibrations, the new scattering channel related to inelastic excitation of the local mode opens \cite{Jaklevic1966,Lambe1968,Zhu2003}.

This paper is organized as follows: In Sec.~\ref{sec:analytics} we present a microscopic  Hamiltonian and provide a brief description of the methods used in our study. The results of the numerical simulations are presented in Sec.~\ref{sec:numerics}. Finally, we summarize our findings in Sec.~\ref{sec:conclusions}.

\begin{figure}[!h]
\centering{}\includegraphics[scale=0.5]{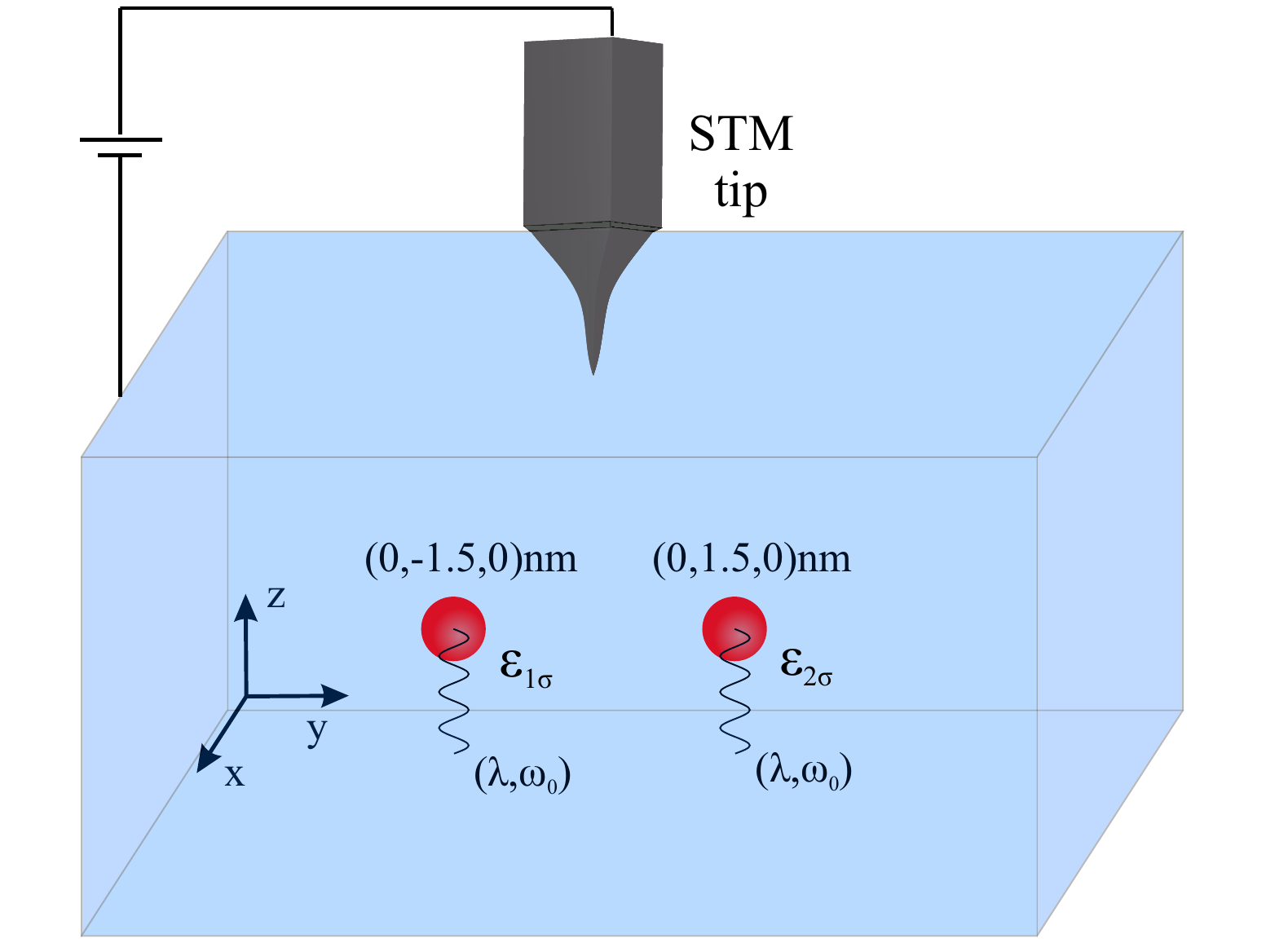}
\protect\caption{\label{fig:Pic1} (Color online) Schematic representation of the model system: two impurities, positioned at $\mathbf{R}_{1}$ and $\mathbf{R}_{2}$ with energy levels $\varepsilon_{1\sigma}$ and $\varepsilon_{2\sigma}$, are embedded into a three-dimensional Dirac semimetal. We account for the vibrational degrees by coupling a phonon mode of the coupling strength $\lambda$ and energy $\omega_{0}$ to each of the impurities.}
\end{figure}
\section{Analytical results}\label{sec:analytics}

In this section we provide the microscopic Hamiltonian and introduce the analytical technique used throughout the paper. For simplicity, we consider the coupling to a single-phonon mode present in the system and treat electron-phonon interaction by means of Lang-Firsov transformation. We further utilize the equation-of-motion (EOM) approach \cite{Marques2017,Haug1996} to evaluate the local density of states (LDOS) and see how the presence of the vibrational degrees of freedom affects the formation of antibonding state (we put $\hbar=1$ throughout the calculations).

\subsection{Model Hamiltonian}\label{sec:hamiltonian}
Consider the geometry schematically shown in the Fig.~\ref{fig:Pic1}: two localized impurities are embedded into 3D-DSM, and each of them is supposed to possess vibrational degree of freedom. The Hamiltonian of such a system may be represented as
\begin{eqnarray}
\mathcal{H}& = & \mathcal{H}_{\textrm{0}}+\mathcal{H}_{\textrm{d}}+\mathcal{H}_{\textrm{hyb}}+\mathcal{H}_{\textrm{e-ph}}+\mathcal{H}_{\textrm{ph}}.\label{eq:H_T}
\end{eqnarray}
In Eq.~(\ref{eq:H_T}) the first term corresponds to the low energy effective Hamiltonian of 3D-DSM,
\begin{eqnarray}
\mathcal{H}_0 & = & \sum_{\mathbf{k}\sigma}\sum_{\tau=\pm}\tau v_{F}(\mathbf{k}\cdot\boldsymbol{\sigma})c_{\mathbf{k}\sigma\tau}^{\dagger}c_{\mathbf{k}\sigma\tau}\label{eq:H_0},
\end{eqnarray}
where the three-dimensional wave vector $\mathbf{k}=(k_{x},k_{y},k_{z})$ specifies electronic states, $\boldsymbol{\sigma}$ stands for a vector of Pauli matrices, and $v_{F}$ is the Fermi velocity at the Dirac point. Moreover, $c_{\mathbf{k}\sigma\tau}(c_{\mathbf{k}\sigma\tau}^{\dagger})$ stands for the annihilation (creation) operator of an electron with spin
$\sigma$ and chirality $\tau$. The second term in (\ref{eq:H_T}) describes a pair of impurities, modeled by the Hubbard-type Hamiltonian
\begin{eqnarray}
\mathcal{H}_d & = & \sum_{j\sigma}\varepsilon_{j\sigma}d_{j\sigma}^{\dagger}d_{j\sigma}+\sum_{j}U_{j}n_{j\uparrow}n_{j\downarrow},\label{eq:H_d}
\end{eqnarray}
with single-particle energy $\varepsilon_{j\sigma}$ and on-site Coulomb repulsion $U_{j}$, whereas $d_{j\sigma}^{\dagger}$ and $d_{j\sigma}$ label creation and annihilation operators respectively, $n_{j\sigma}=d_{j\sigma}^{\dagger}d_{j\sigma}$ counts the number of electrons with spin projection $\sigma$ at site $j$, and $n_j=n_{j\uparrow}+n_{j\downarrow}$. The third term in (\ref{eq:H_T}),
\begin{eqnarray}
\mathcal{H}_{\mathrm{hyb}} & = & \sum_{j\mathbf{k}\sigma}\sum\limits_{\tau=\pm}(V_{j\mathbf{k}}d_{j\sigma}^{\dagger}c_{\mathbf{k}\sigma\tau}+V_{j\mathbf{k}}^{*}c_{\mathbf{k}\sigma\tau}^\dagger d_{j\sigma}),\label{eq:H_V}
\end{eqnarray}
represents the hybridization between the electronic states of impurities and conduction electrons with the tunneling matrix element $V_{j\mathbf{k}}=v_{0}e^{i\mathbf{k}\cdot\mathbf{R}_{j}}/\sqrt{N}$, the amplitude $v_{0}$ is chosen the same for both impurities positioned at $\mathbf{R}_{j}$ ($j=1,2$), $N$ is the normalization factor yielding the total number of states. This element well describes the electronic scattering by impurities and neglects the exponential decay towards the STM tip. It means that our findings would be just attenuated at the surface, without loss of generality. The fourth term in (\ref{eq:H_T}) corresponds to the electron-phonon interaction, thus accounting for the possibility of impurity vibrations
\begin{eqnarray}
\mathcal{H}_{\textrm{e-ph}} & = & \lambda\sum_j n_j(a_{j}^{\dagger}+a_{j}),\label{eq:H_eph}
\end{eqnarray}
where $\lambda$ is the electron-phonon coupling strength, while $a_j\,(a_j^\dagger)$ stands for annihilation (creation) operator of bosonic mode at $j$th site. The last term in (\ref{eq:H_T}),
\begin{eqnarray}
\mathcal{H}_{\mathrm{ph}} & = & \omega_{0}\sum_{j}a_{j}^{\dagger}a_{j},\label{eq:H_ph}
\end{eqnarray}
corresponds to the free phonons. We restrict our analysis to a single mode with the frequency $\omega_0$.

\subsection{Lang-Firsov transformation}\label{sec:transformation}
The Hamiltonian (\ref{eq:H_T}) resembles that of two-site Anderson-Holstein model, interacting with a single vibron mode, that can be rigorously approached using Lang-Firsov canonical transformation \cite{Lang1963,Mahan2000}. The latter eliminates the electron-phonon coupling ($\mathcal{H}_{\textrm{e-ph}}$) and leads to the renormalization of impurity elements ($\mathcal{H}_{\textrm{d}}, \mathcal{H}_{\textrm{hyb}}$). The transformed Hamiltonian $\tilde{\mathcal{H}}=S\mathcal{H}S^{-1}$ reads as
\begin{subequations}
\begin{equation}\label{eq:T_H}
\tilde{\mathcal{H}}=\mathcal{H}_0+\tilde{\mathcal{H}}_d+\tilde{\mathcal{H}}_\mathrm{hyb}+\mathcal{H}_\mathrm{ph},
\end{equation}
\begin{equation}
\tilde{\mathcal{H}}_d=\sum\limits_{j\sigma}\tilde{\varepsilon}_{j\sigma}n_{j\sigma}+\sum\limits_j\tilde{U}_jn_{j\uparrow}n_{j\downarrow},
\end{equation}
\begin{equation}
\tilde{\mathcal{H}}_\mathrm{hyb}=\sum\limits_{j\mathbf{k}\sigma}\sum\limits_{\tau=\pm}(V_{j\mathbf{k}} \tilde{d}_{j\sigma}^\dagger c_{\mathbf{k}\sigma\tau}+V_{j\mathbf{k}}^{*} c_{\mathbf{k}\sigma\tau}^\dagger \tilde{d}_{j\sigma}),
\end{equation}
\end{subequations}

\noindent where $S=\exp[\sum_jx_jn_j]$ is anti-Hermitian generator with $x_j=\lambda(a_j-a_j^\dagger)/\omega_0$, while $\tilde{\varepsilon}_{j\sigma}=\varepsilon_{j\sigma}-\frac{\lambda^{2}}{\omega_{0}}$ and $\tilde{U}_{j}=U_{j}-\frac{2\lambda^{2}}{\omega_{0}}$ represent the renormalization of the particle energy and on-site Coulomb repulsion due to their coupling with the $j-$th phonon. The absence of an explicit electron-phonon coupling term in Eq.~(\ref{eq:T_H}) is justified by the renormalization $\tilde{d}_{j\sigma}=d_{j\sigma}\hat{X}_j$ in the  hybridization term by the phonon shift generator $\hat{X}_j=e^{x_j}$.
The price one has to pay for that type of transformation is the necessity to work with polaron-type qusiparticles, i.e., electrons surrounded by phonon clouds. The effects of the polaronic interactions in the molecule ground state
can be analyzed via probing LDOS of the system, e.g., with the application of STM, defined as
\begin{equation}
\mathrm{LDOS}\,(\varepsilon,\mathbf{R}_{m})=-\frac{1}{\pi}\sum_{\sigma}\mathrm{Im}\,[G^R_{\sigma}(\varepsilon,\mathbf{R}_{m})],\label{eq:LDOS}
\end{equation}
where
\begin{equation}
G^R_{\sigma}(\varepsilon,\mathbf{R}_{m})=\frac{1}{N}\sum_{\mathbf{k}\mathbf{k}'}\sum_{\tau\tau'}e^{-i(\mathbf{k}-\mathbf{k}')\cdot\mathbf{R}_{m}}G^{ R}_{\mathbf{k}\sigma\tau\vert\mathbf{k}'\sigma\tau'}(\varepsilon),\label{eq:Main_GF}
\end{equation}
is the retarded Green's function of the system in the energy domain at the STM tip position $\mathbf{R}_{m}$. We further proceed our analysis with EOM approach to the Green's functions of conduction electrons in time domain
\begin{equation}
G^R_{\mathbf{k}\sigma\tau\vert\mathbf{k}'\sigma\tau'}(t)=-i\theta(t)\langle\{c_{\mathbf{k}\sigma\tau}(t),c^\dagger_{\mathbf{k}'\sigma\tau'}(0)\}\rangle_{\tilde{\mathcal{H}}},
\end{equation}
where the angular brackets $\langle\ldots\rangle_{\tilde{\mathcal{H}}}$ denote the thermal averaging with respect to the transformed Hamiltonian (\ref{eq:T_H}). In a similar way, we define Green's functions of $d$ electrons
\begin{align}\nonumber
\mathcal{G}_{j\sigma\vert j'\sigma}^R(t) & = -i\theta(t)\langle\{\tilde{d}_{j\sigma}(t),\tilde{d}_{j'\sigma}^\dagger(0)\}\rangle_{\tilde{\mathcal{H}}} \\
& = -i\theta(t)\langle\{d_{j\sigma}(t),d_{j'\sigma}^\dagger(0)\}\rangle_{\mathcal{H}}.\label{green}
\end{align}

\subsection{Equation-of-motion approach}\label{sec:eom}
In general, the EOM formalism implies calculation of time derivatives of the Green's functions of the system. If the system Hamiltonian contains the interaction terms, these derivatives lead to higher-order Green's functions which at some point should be truncated to close the system of differential equations.

Applying standard EOM techniques to $G^R_{\mathbf{k}\sigma\tau\vert\mathbf{k}'\sigma\tau'}(\varepsilon)$ and performing summation over $\tau$ and $\tau'$ described in Eq.~(\ref{eq:Main_GF}), we derive
\begin{align}\nonumber
\sum\limits_{\tau\tau'}G^R_{\mathbf{k}\sigma\tau\vert\mathbf{k}'\sigma\tau'}(\varepsilon)\\ = f_\mathbf{k}\delta_\mathbf{kk'}+f_\mathbf{k}f_{\mathbf{k}'}\sum\limits_{jj'}V_{j\mathbf{k}}^{*} \,\mathcal{G}_{j\sigma\vert j'\sigma}^R(\varepsilon)\,V_{j'\mathbf{k}'}, \label{eq:PR SIGMA}
\end{align}
where $f_\mathbf{k}=2\varepsilon/(\varepsilon^2-v_F^2k^2)$.
Thereby, Eq.~(\ref{eq:LDOS}) gives
\begin{equation}\label{eq:LDos}
\mathrm{LDOS}(\varepsilon)=\rho_0(\varepsilon)+\sum\limits_{\sigma}\sum\limits_{jj'}\delta\rho^\sigma_{jj'}(\varepsilon),
\end{equation}
where $\rho_{0}(\varepsilon)\equiv-\frac{1}{\pi N}\textrm{Im}\{ \sum_{\mathbf{k}}f_\mathbf{k}\}=V\varepsilon^2/(N\pi^2\hbar^3v_F^3)=3\varepsilon^2/D^3$ stands for the bare 3D-DSM density of states (here, $V$ is the volume occupied by the system and we defined the half-bandwidth of the system $D$), while the contribution stemming from the scattering of conduction electrons off localized impurities may be represented as
\begin{equation}\label{de-ldos}
\delta\rho_{jj'}^\sigma(\varepsilon)=-\frac{1}{\pi v_0^2}\mathrm{Im}\left[\Sigma_{mj}(\varepsilon)\mathcal{G}^R_{j\sigma\vert j'\sigma}(\varepsilon)\Sigma_{j'm}(\varepsilon)\right].
\end{equation}
Note that Eq.~(\ref{de-ldos}) corresponds to the electronic waves scattered by the localized impurities directly into the host semimetal for $j'=j$ and the waves scattered between the impurities via host if $j'$ is opposite to $j$, i.e., $j'=\bar{j}$. Remarkably, the self-energy determined in (\ref{de-ldos}),
\begin{equation}
\Sigma_{mj}(\varepsilon)=\frac{v_{0}^{2}}{N}\sum_{\mathbf{k}}f_\mathbf{k}\,e^{i\mathbf{k}\cdot(\mathbf{R}_{m}-\mathbf{R}_{j})}\label{eq:SE_Rmj},
\end{equation}
is not affected by the vibrations given that just localized phonon modes are considered in our model. However, the Green's function of the localized levels (\ref{green})
is renormalized by the phonon shift generator, thus being responsible for the coupling between the electron and phonon subsystems.

To get a further analytical insight, we decouple the electron Green's function of Eq.~(\ref{green}) using analog of the Born-Oppenheimer approximation \cite{Mahan2000,Zhu2003}, thus treating electron and phonon degrees of freedom separately,
\begin{subequations}
\begin{equation}\label{eq:G_dX-1}
\mathcal{G}_{j\sigma\vert j'\sigma}^R(t)\thickapprox g^R_{j\sigma\vert j'\sigma}(t)\langle\hat{X}_j(t)\hat{X}_{j'}^{\dagger}(0)\rangle _{\bar{\mathcal{H}}_{\mathrm{ph}}},
\end{equation}
\begin{equation}
g^R_{j\sigma\vert j'\sigma}(t)=-i\theta(t)\langle\{d_{j\sigma}(t),d_{j'\sigma}^{\dagger}(0)\}\rangle _{\bar{\mathcal{H}}_{\mathrm{el}}},
\end{equation}
\end{subequations}
where $\hat{X}_j(t)=e^{i\tilde{\mathcal{H}}_{\mathrm{ph}}t}\hat{X}_je^{-i\tilde{\mathcal{H}}_{\mathrm{ph}}t}$. Interestingly, averaging of phonon shift operators dresses diagonal ($j'=j$) and off-diagonal ($j'\neq j$) components of Eq.~(\ref{eq:G_dX-1}) in a different way, namely,
\begin{eqnarray}\nonumber
\mathcal{G}_{j\sigma\vert j\sigma}^R(\varepsilon)=e^{-\beta\coth z} \\ \times\sum\limits_{n=-\infty}^\infty g^R_{j\sigma\vert j\sigma}(\varepsilon-n\omega_0)\,I_n(2\beta\,\mathrm{csch}\,z)\,e^{nz},\label{eq:Re_jj}
\end{eqnarray}
and another for $j'=\bar{j}\neq j$,
\begin{equation}
\mathcal{G}_{j\sigma\vert \bar{j}\sigma}^R(\varepsilon)=g_{j\sigma\vert \bar{j}\sigma}^R(\varepsilon)\,e^{-\beta\coth z},
\end{equation}
$z=\omega_0/(2kT)$, $\beta=\lambda^{2}/\omega_{0}^{2}$, while $I_{n}(y)$ stands for the modified $n$th-order Bessel function for complex arguments, $\coth y=(e^y+e^{-y})/(e^y-e^{-y})$ and $\mathrm{csch}\,y=2/(e^y-e^{-y})$ denote hyperbolic cotangent and hyperbolic cosecant, respectively. The pure electronic Green's function $g_{j\sigma\vert j'\sigma}^R(\varepsilon)$ is evaluated by means of its EOM with respect only to the electronic part of the Hamiltonian described in Eq.~(\ref{eq:T_H}), leading to
\begin{align}\nonumber
[\varepsilon-\tilde{\varepsilon}_{j\sigma}-&\tilde{\Sigma}_{jj}(\varepsilon)]\,g^R_{j\sigma\vert j'\sigma}(\varepsilon) =\delta_{jj'} \\ &+\tilde{U}_{j}\,\mathcal{G}_{j\sigma\bar{\sigma}\vert j'\sigma}^{(4)}(\varepsilon)+\tilde{\Sigma}_{j\bar{j}}(\varepsilon)\,g^R_{\bar{j}\sigma\vert j'\sigma}(\varepsilon),\label{purel}
\end{align}
where renormalized self-energies $\tilde{\Sigma}_{jj}=\langle \hat{X}_j\hat{X}_j^{\dagger}\rangle_{\tilde{\mathcal{H}}_\mathrm{ph}}\Sigma_{jj}$ and $\tilde{\Sigma}_{j\bar{j}}=\langle \hat{X}_j\hat{X}_{\bar{j}}^\dagger\rangle_{\tilde{\mathcal{H}}_\mathrm{ph}}\Sigma_{j\bar{j}}$. In Eq.~(\ref{purel}) we have also introduced the two-particle Green's function $\mathcal{G}^{(4)}_{j\sigma\bar{\sigma}\vert j'\sigma}(t)=-\theta(t)\langle\{d_{j\sigma}(t)n_{j\bar{\sigma}}(t),d_{j'\sigma}^\dagger(0)\}\rangle$ with $\bar{\sigma}=-\sigma$ that is known to be responsible for generating an infinite set of high-order Green's functions. To truncate the infinite hierarchy of such obtained equations, we make use of the Hubbard I decoupling scheme \cite{Marques2017,Hubbard1964}. This approach provides a reliable description of impurities away from the Kondo regime in metals. Such a phenomenon appears as a sharp peak at the Fermi level in the impurities density of states for low-temperatures regimes ($T\ll T_{K}$, where $T_{K}$ is the Kondo temperature of the system). Accounting for the fact that in 3D-DSM $\rho_{0}(\varepsilon_{F})=0$, we can safely apply the Hubbard I framework in the current system even with $T\ll T_{K}$. After straightforward algebra, we derive
\begin{align}\nonumber
[\varepsilon-\tilde{\varepsilon}_{j\sigma}-\tilde{U}_{j}-\sum\limits_\mathbf{k}f_\mathbf{k}\vert V_{j\mathbf{k}}\vert^2\langle \hat{X}_j\hat{X}_j^{\dagger}\rangle_{\tilde{\mathcal{H}}_\mathrm{ph}}]\,\mathcal{G}^{(4)}_{j\sigma\bar{\sigma}\vert j'\sigma}(\varepsilon) \\ =\left\langle n_{j\bar{\sigma}}\right\rangle[\delta_{jj'}+\sum_{\mathbf{k}}f_\mathbf{k}V_{j\mathbf{k}}^{*} V_{\bar{j}\mathbf{k}}\langle \hat{X}_j\hat{X}_{\bar{j}}^\dagger\rangle_{\tilde{\mathcal{H}}_\mathrm{ph}}g_{\bar{j}\sigma\vert j'\sigma}^R(\varepsilon)].\label{eq:hb-1}
\end{align}

\noindent The expression (\ref{eq:hb-1}) can be closed upon identifying the fermionic occupation number according to
\begin{eqnarray}
\left\langle n_{j\sigma}\right\rangle  & = & -\frac{1}{\pi}\int_{-D}^{D}n_{F}(\varepsilon)\,\mathrm{Im}[g_{j\sigma\vert j\sigma}^R(\varepsilon)]d\varepsilon, \label{eq:Occ}
\end{eqnarray}
with $n_{F}(\varepsilon)=1/(e^{\varepsilon/kT}+1)$ being the Fermi-Dirac distribution. Finally, combining Eqs. (\ref{eq:Re_jj})--(\ref{eq:hb-1}),
we find the diagonal components of Green's function as
\begin{equation}
\mathcal{G}_{j\sigma\vert j\sigma}^R(\varepsilon)=\sum_{n=-\infty}^{\infty}\frac{\lambda_{j}^{\bar{\sigma}}I_{n}(2\beta\,\mathrm{csch}\,z)\,e^{nz-\beta\coth z}}{\varepsilon-n\omega_{0}-\tilde{\varepsilon}_{j\sigma}-\varLambda_{jj}^{\sigma}(\varepsilon-n\omega_{0})},\label{eq:DiagGf}
\end{equation}
whereas off-diagonal components of Green's function may be determined from
\begin{align}
\mathcal{G}_{j\sigma\vert\bar{j}\sigma}^R(\varepsilon)=\frac{\tilde{\Sigma}_{j\bar{j}}(\varepsilon)}{\varepsilon-\tilde{\varepsilon}_{j\sigma}-\tilde{\Sigma}(\varepsilon)}\frac{\lambda_{j}^{\bar{\sigma}}\lambda_{\bar{j}}^{\bar{\sigma}}e^{-\beta\coth z}}{\varepsilon-\tilde{\varepsilon}_{j\sigma}-\varLambda_{jj}^{\sigma}(\varepsilon)},\label{eq:MixedGf}
\end{align}
provided that
\begin{equation}
\lambda_{j}^{\bar{\sigma}}=1+\frac{\tilde{U}_{j}\left\langle n_{j\bar{\sigma}}\right\rangle}{\varepsilon-\tilde{\varepsilon}_{j\sigma}-\tilde{U}_{j}-\tilde{\Sigma}_{0}(\varepsilon)},
\end{equation}
and
\begin{equation}\label{se}
\varLambda_{jj}^{\sigma}(\varepsilon)=\tilde{\Sigma}_{0}(\varepsilon)+\frac{\tilde{\Sigma}_{j\bar{j}}(\varepsilon)\tilde{\Sigma}_{j\bar{j}}(\varepsilon)}{\varepsilon-\tilde{\varepsilon}_{\bar{j}\sigma}-\tilde{\Sigma}_{0}(\varepsilon)}.
\end{equation}
It should be pointed out that although the Born-Oppenheimer decoupling apparently disregards any correlation between electron and phonon degrees of freedom, both local $(\Sigma_{0})$ and non-local $(\Sigma_{j\bar{j}})$ electronic self-energies, defined in Eq.~(\ref{eq:SE_Rmj}), are renormalized by the phonon field, so that $\tilde{\Sigma}_{0}(\varepsilon)=\Sigma_{0}(\varepsilon)\,e^{-\beta\coth z}$ and $\tilde{\Sigma}_{j\bar{j}}(\varepsilon)=\Sigma_{j\bar{j}}(\varepsilon)\,e^{-\beta\coth z}$, respectively. After summating over all available states in the $\mathbf{k}$ space and introducing the energy cutoff $D$ as the half-bandwidth of the
3D-DSM, the bare self-energy may be written as follows:
\begin{align}
\Sigma_{0}^R(\varepsilon)=\frac{3v_{0}^{2}}{D^{3}}\left(\varepsilon^{2}\ln\left|\frac{D+\varepsilon}{D-\varepsilon}\right|-2D^{2}-i\pi\varepsilon^{2}\right),
\end{align}
for $\mathbf{R}_{jj}=0$ and
\begin{align}
\Sigma_{j\bar{j}}^R(\varepsilon)=-\frac{3 \pi v_{0}^{2}}{D^{3}}\frac{\varepsilon \hbar v_{F}}{|\mathbf{R}_{j\bar{j}}|}\exp\left(i\frac{\varepsilon|\mathbf{R}_{j\bar{j}}|}{\hbar v_{F}}\right)\label{eq:SE Rmj-1},
\end{align}

\noindent for $\mathbf{R}_{j\bar{j}}=\mathbf{R}_{j}-\mathbf{R}_{\bar{j}}$. We have also established that the presence of vibrations results in the diagonal components of the Green's function of the system (\ref{green}) being dressed, thus creating an infinite set of polaronic states. This entire set of states is modulated by the exponential factor which depends on the ratio $\lambda/\omega_{0}$ and is quite sensitive to the temperature $T$. The latter suggests that tuning these quantities determines how many polaronic states are involved in the formation of the ground state.

\section{Numerical results}\label{sec:numerics}
In order to explore the role of polaronic interactions
in the formation of the molecular state of two identical impurities placed at $R_{1,2}=(0,\pm1.5,0)$ nm with energy levels $\varepsilon_{j\sigma}=-0.1D$, on-site Coulomb repulsion $U_j=0.2D$ and hybridization strength $v_0=0.2D$, we consider a phonon mode with energy $\omega_0=0.01D$ coupled to each impurity as sketched in Fig.~(\ref{fig:Pic1}). Noteworthy, it was already shown in Ref.~\cite{Marques2017} that in the absence of vibrations, the emergence of antibonding ground state is independent on the choice of the electronic parameters ($\varepsilon_{j\sigma}$,$U_{j}$,$v_{0}$) as long as these quantities do not create resonant levels very far from the Dirac point, where the linear dispersion relation approximation does not hold anymore. For sake of feasibility, we follow $D\approx0.2\,\text{\text{{eV}}}$ with $\hbar v_{F}\approx5\:eV\text{\AA}$ for Cd$_3$As$_2$ \cite{Borisenko2014,Neupane2014}. The diagonal components of Green's function described in Eq.~(\ref{eq:DiagGf}) generates an ensemble of polaronic sidebands with energies $\tilde{\varepsilon}_{j\sigma}$ owing to the presence of the poles at $\varepsilon=n\omega_{0}$, so that high phonon energies may give rise to the resonance levels near the band threshold. Meanwhile, within Born-Oppenheimer approximation, these levels appear only around the Fermi energy which, according to Ref.\cite{Hewson1979}, represents a suitable description for local polaronic interactions in the weak coupling limit ($\omega_{0},\lambda\ll v_{0}$). Furthermore, if we consider $\omega_{0}$ in a strong coupling limit ($\omega_{0},\lambda\gg v_{0}$), for instance, the diagonal Green's function should be renormalized just by $e^{-g\coth z}$  instead of the summation found in Eq.~(\ref{eq:DiagGf}) \cite{Hewson1979}. However, no qualitative change in the nature of the ground state shows up in this scenario.
\begin{figure}[!h]
\centering{}\includegraphics[scale=0.43]{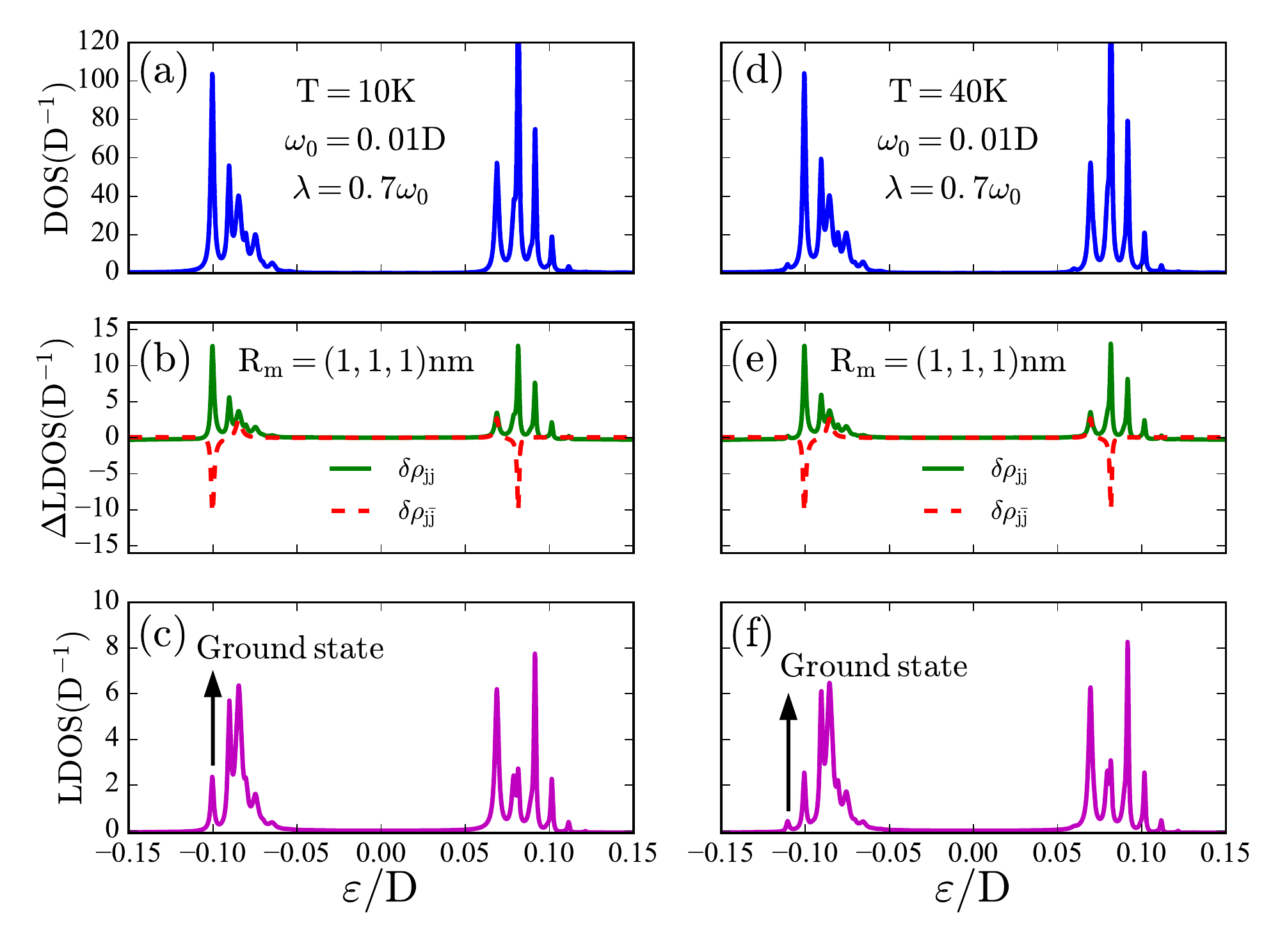}
\protect\caption{\label{fig:Pic2} (Color online) Electronic properties of the system in the valence band with $\varepsilon_{j\sigma}=-0.10D$, $U_{j}=0.20D$, $v_{0}=0.20D$,
$\omega_{0}=0.01D$, and $\lambda=0.7\omega_{0}$ for $T=10$ K (left
panels) and $T=40$ K (right panels). Impurities density of
states presenting the electronic and polaronic levels (a) and (d).  The profiles of $\delta\rho_{jj'}(\omega)$ at $\mathbf{R}_{m}=(1,1,1)\text{{ nm}}$ for $j'=j$ (solid green curve) and $j'=\bar{j}$ (dashed red curve)
revealing the interferences between diagonal and mixed electronic
waves scattering (b) and (e). Total LDOS (c), described in Eq.~(\ref{eq:LDos}), displaying a pure electronic ground
state formed by a destructive interference shown in panel (b). Total LDOS revealing the emergence of a lowest energy polaronic mode due to the thermal excitation (f). The insets display the entire spectrum of each panel, i.e., valence and conduction bands.}
\end{figure}

As a pair of identical impurities is considered, their densities of states, defined by $\mathrm{LDOS}_{jj}(\varepsilon)=-\frac{1}{\pi}\mathrm{Im}[\sum_{\sigma}\mathcal{G}_{j\sigma\vert j\sigma}^R(\varepsilon)]$, exhibit exactly the same behavior. In Fig.~\ref{fig:Pic2}(a) these densities of states for $\lambda=0.7\omega_{0}$ and $T=10$ K present sidebands in the valence band. Close inspection of Fig.~\ref{fig:Pic2}(b), which displays the induced local density of states [$\delta\rho_{jj'}(\varepsilon)$] at $\mathbf{R}_{m}=(1,1,1)$ nm, reveals that the mixed term indicates the energetic position of the pure electronic states since there is no polaronic interaction in the mixed Green's function described by Eq.~(\ref{eq:MixedGf}). Together with them, we also find four peaks in the diagonal Green's function for $n=0$. Thereby, both the destructive and constructive interference at these energies create pure electronic states, thus indicating that all other resonances correspond to polaronic modes. We highlight that despite the low temperature ($T=10$ K), a set polaronic modes emerges just because the electronic tunneling between impurity and host, described by the pure electronic state at $\varepsilon=-0.09D$, excites the phonon degree of freedom. Thereby, all polaronic modes exclusively excited by the electron tunneling are located at highest energetic positions in comparison with state at $\varepsilon=-0.10D$. Upon increase of the temperature to $T=40$ K, some thermally excited phonons are absorbed in process of the electronic scattering, giving rise to additional modes in the density of states as can be seen in Fig.~\ref{fig:Pic2}(d). As an aftermath, the LDOS displayed in Fig.~\ref{fig:Pic2}(f) exhibits a different ground state \cite{ground} as compared to the LDOS presented in Fig.~\ref{fig:Pic2}(c). In summary, for $T=10$ K the ground state is created by the destructive interference between pure electronic states, whereas for $T=40$ K the constructive interference [$\delta\rho_{11}(\varepsilon)+\delta\rho_{22}(\varepsilon)$] of the polaronic modes at $\varepsilon=-0.11D$ creates a new ground state. We call attention that in Figs.~\ref{fig:Pic2}(c) and (f) the background $\rho_{0} \sim \varepsilon ^{2}$ is not noticed due to the pronounced induced LDOS amplitude $\delta\rho_{jj'}$. In the latter, the ratio $\hbar v_{F}/D|\mathbf{R}_{mj}|$ from Eq.~(\ref{eq:SE Rmj-1}) modulates such amplitude as can be seen in Eq.~(\ref{de-ldos}), while in the former $v_{F}$ enters inversely. It means that one should properly choose the separation between the impurities and Fermi velocity, which is material dependent, to notice the background scaling. Moreover, as we considered the weak coupling regime, the peak's broadening is ruled by the Anderson parameter $ \Gamma (\varepsilon)=\pi v_{0}^{2} \rho(\varepsilon)$ with $\rho(\varepsilon)\propto\varepsilon ^{2}$. Hence, for energies near the Fermi level ($\varepsilon/D \ll 1$), the broadening is small enough to form sharp peaks.

\begin{figure}[!h]
\centering{}\includegraphics[scale=0.43]{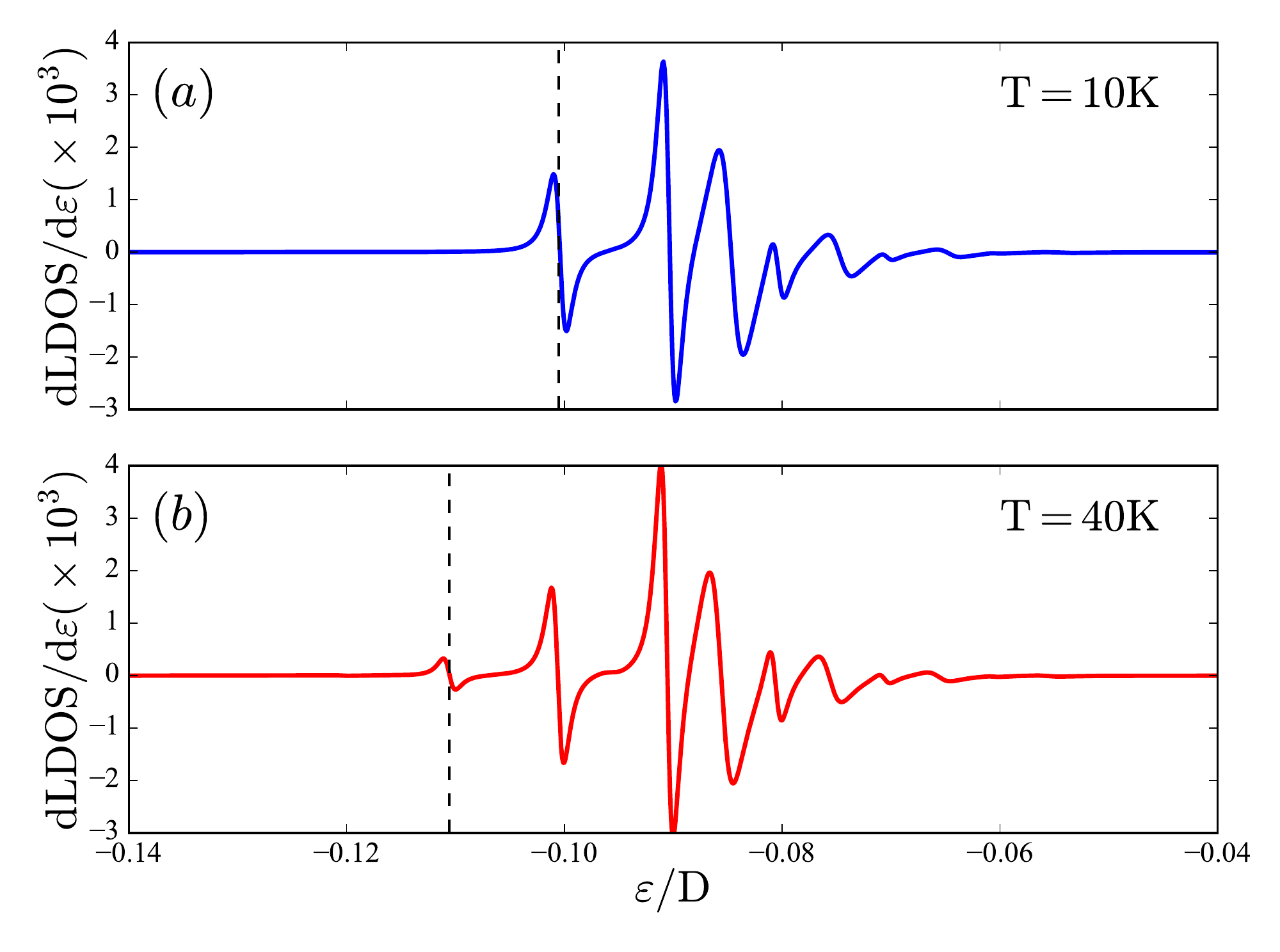}
\protect\caption{\label{fig:Pic2b}(Color online) The $d\mathrm{LDOS}(\varepsilon)/d\varepsilon=0$ profile on the valence bands with the parameters: $\varepsilon_{j\sigma}=-0.10D$, $U_{j}=0.20D$, $v_{0}=0.20D$, $\omega_{0}=0.01D$ and $\lambda=0.7\omega_{0}$ for $T=10K$ (a) and $T=40K$ (b). The vertical dashed lines mark the energetic position of the ground states.
}\label{fig:fig4}
\end{figure}
The detection of inelastic features is usually performed by looking at $d^{2}I/dV^{2} \propto d\mathrm{LDOS}(\varepsilon)/d\varepsilon$ \cite{Jaklevic1966,Lambe1968,Zhu2003}. However, as the bonding ground state is assisted by vibrations, it emerges in the DOS as just a peak. Consequently, such state is of resonant type as well as the corresponding antibonding state, which is of pure electronic nature. The direct implication is that both signals reveal a similar profile as shown in Fig.~\ref{fig:fig4}, wherein the former stems from resonant inelastic electron-phonon scattering. If temperature is further increased, other assisted phonon sidebands would emerge configuring new ground states, but all of them with bonding characteristics once that the unique antibonding state comes from an indirect tunneling that is decoupled from the local phonon perturbation, thus preventing any new bonding-antibonding crossover.

\begin{figure}[!h]
\centering{}\includegraphics[scale=0.43]{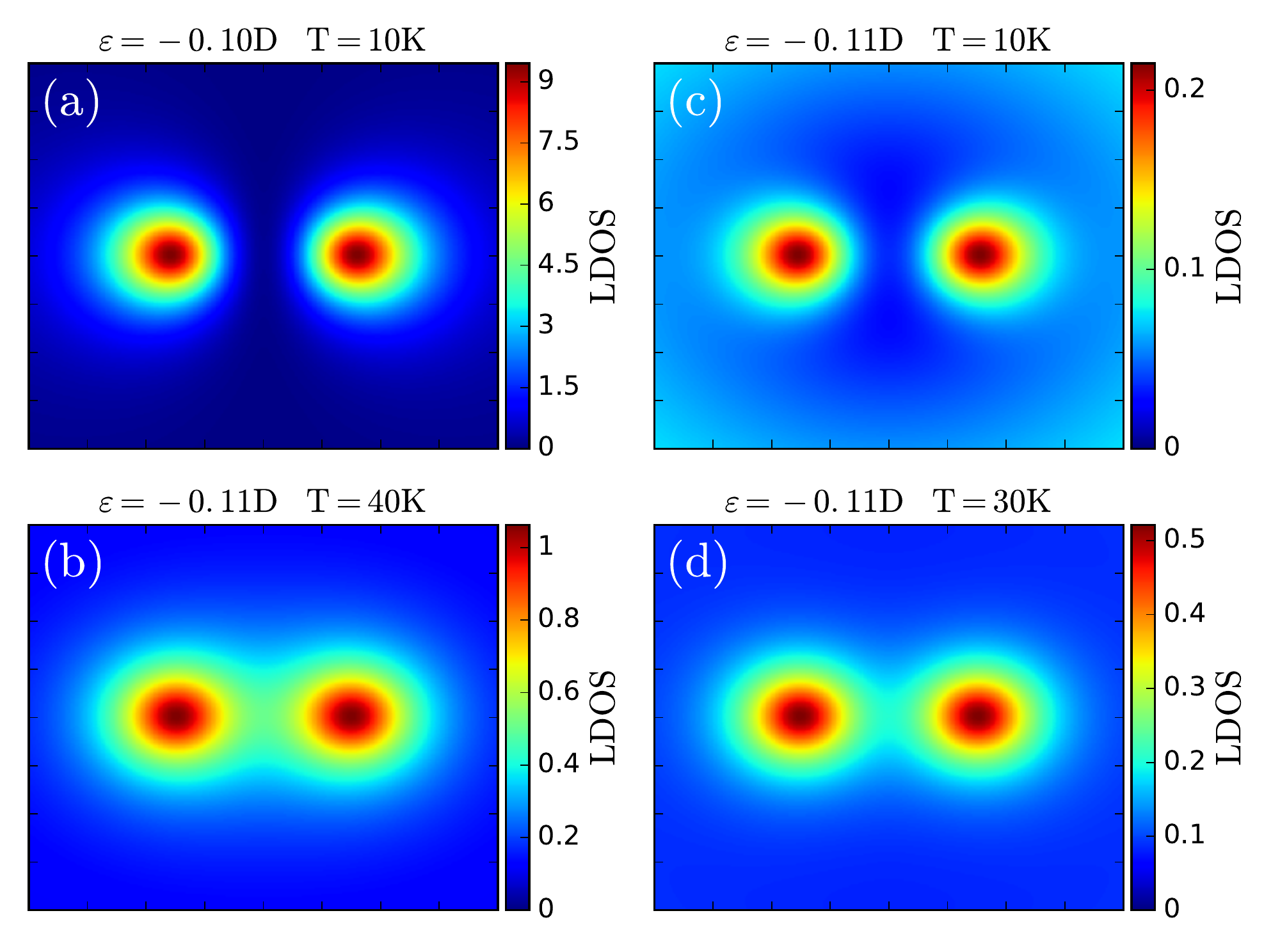}
\protect\caption{\label{fig:Pic3} (Color online) LDOS topography of the surface of
3D-DSM [$\mathbf{R}_{m}=(x,y,1)\text{{ nm}}$ plane]. For $T=10$ K (a), at the ground state energy $\varepsilon=-0.10D$, an antibonding profile emerges due to the destructive interference between the waves scattered by the impurities. For higher temperature $T=40$ K (b), the ground state at $\varepsilon=-0.11D$ features bonding characteristics. The orbital characteristics of the state $\varepsilon=-0.11D$ for $T=10$ K (c) and $T=30$ K (d), respectively, revealing smooth crossover in the nature of the ground state.
}
\end{figure}
To better understand the change in the nature of the ground state, a topographic analysis of the LDOS on the surface $\mathbf{R}_{m}=(x,y,1)$ nm is presented in Fig.~\ref{fig:Pic3}. We start by analyzing the surface LDOS for the two energies of interest, $\varepsilon=-0.10D$ and $\varepsilon=-0.11D$ which correspond to the ground states found in Fig.~\ref{fig:Pic2} for $T=10$ and $40$ K, respectively. In Fig.~\ref{fig:Pic3}(a), an antibonding profile characterizes the ground state of the molecule. However, the thermally excited polaronic state which emerges for high temperatures, shown in Fig.~\ref{fig:Pic2}(f), creates a new ground state of a bonding character as can be seen in Fig.~\ref{fig:Pic3}(b). To comprehend this transition, we should look at the surface LDOS at $\varepsilon=-0.11D$ for $T=10$ K displayed in Fig.~\ref{fig:Pic3}(c). One can notice highly attenuated LDOS signal presenting an antibonding characteristic, pointing out that this is just the spread of the ground state found at $\varepsilon=-0.10D$, as shown in Fig.~\ref{fig:Pic3}(a). This means that indeed there is no thermally excited state for $T=10$ K. In contrast, for $T=30$ K displayed in Fig.~\ref{fig:Pic3}(d), the LDOS signal is enhanced by thermally excited polarons. These results show, thus, that the transition between ground states has crossover character.

The formation of an antibonding ground state was first discovered in a system of coupled quantum dots \cite{Doty2009,Yakimov2011}. It is manifested as negative electronic hopping between the dots mediated by spin-orbit coupling. In the meantime, it turns out that a similar effect takes place in the model in question: Indeed, Friedel-type oscillations of electron density well inside 3D-DSM results in the effective hopping term $\varLambda_{jj}^{\sigma}(\varepsilon)$ of Eq.~(\ref{se}) between the impurities being negative. In quantum dot systems, the antibonding ground state is found at zero temperature. Hence, the molecule is always found in its ground state due to the maximum probability given by the statistical weight of Boltzmann factor. For finite temperatures, however, there is a probability to find the molecule out of the ground state. As the molecular spectrum changes with temperature due to the phonon degrees of freedom\cite{ground}, it is not straightforward to determine this probability as a function of temperature. Despite this, for each temperature the ground-state probability can be determined by  $p_{gs}=\frac{1}{Z}e^{-\beta\varepsilon_{gs}}$, where $Z$ is the partition function. Hence, for $T=10$ K the probability to find the molecule in its ground state (antibonding) is $\sim1$, exhibiting that for such temperature the molecule is always in its lowest-energy state. While that for $T=40$ K, the molecule presents $p_{gs} \sim 0.943$ of being in its new ground state (bonding, due to the emergence of the new polaronic level) and with $p_{1^{st}} \sim 0.053$  of being in first excited state (antibonding, which was the former ground state).

\section{Conclusions and outlook}\label{sec:conclusions}

In summary, we theoretically investigated the role of the polaronic interactions in formation of the molecular ground state for two atoms placed inside a 3D-DSM. It was demonstrated that the increasing of the temperature favors the emergence of polaronic modes at lower energetic positions in comparison with pure electronic states that are responsible for the antibonding character of the ground state. As a consequence, the ground state experiences an antibonding-bonding smooth crossover. Our theoretical findings may be probed by means of STM measurements and will trigger further experimental activity in this direction.

\section*{Acknowledgments}

This work was supported by the agencies CNPq (307573/2015-0), CAPES. D.Y. thanks the support from RFBR Project No. 16-32-60040 and Project No. 3.8884.2017/8.9 of the Ministry of Education and Science of the Russian Federation. I.A.S. acknowledges the support from Horizon2020 RISE project CoExAN and RSF(17-12-01581). Y.M. thanks the ITMO University at St. Petersburg for hospitality.


\begin{thebibliography}{10}

\bibitem{Fujikawa2004}K. Fujikawa and H. Suzuki, {\it Path integrals and quantum anomalies} (Clarendon Press, Oxford, 2004).

\bibitem{Weyl1929}H. Weyl, Proc. Natl. Acad. Sci. USA {\bf 15}, 323 (1929).

\bibitem{Bevan1997}T. D. C. Bevan, A. J. Manninen, J. B. Cook, J. R. Hook, H. E. Hall, T. Vachaspati, and G. E. Volovik, Nature {\bf 386}, 689 (1997).

\bibitem{Jiang2012}J.-H. Jiang, Phys. Rev. A {\bf 85}, 033640 (2012).

\bibitem{Halasz2012}G. B. Hal\'asz and L. Balents, Phys. Rev. B {\bf 85}, 035103 (2012).

\bibitem{Manes2012}J. L. Ma\~nes, Phys. Rev. B {\bf 85}, 155118 (2012).

\bibitem{Xu2015}S.-Y. Xu, I. Belopolski, N. Alidoust, M. Neupane, G. Bian, C. Zhang, R. Sankar, G. Chang, Z. Yuan, C.-C. Lee, S.-M. Huang, H. Zheng, J. Ma, D. S. Sanchez, B. K. Wang, A. Bansil, F. Chou, P. P. Shibayev, H. Lin, S. Jia, and M. Z. Hasan, Science {\bf 349}, 613 (2015).

\bibitem{Huang2015}S.-M. Huang, S.-Y. Xu, I. Belopolski, C.-C. Lee, G. Chang, B. K. Wang, N. Alidoust, G. Bian, M. Neupane, C. Zhang, S. Jia, A. Bansil, H. Lin, and M. Z. Hasan, Nat. Commun. {\bf 6}, 7373 (2015).

\bibitem{Armitage2018}N. P. Armitage, E. J. Mele, and A. Vishwanath, Rev. Mod. Phys, (2018).

\bibitem{Murakami2007}S. Murakami, New J. Phys. {\bf 9}, 356 (2007).

\bibitem{Volovik2003}G. E. Volovik, {\it The Universe in a liquid helium droplet} (Clarendon Press, Oxford, 2003).

\bibitem{Wan2011}X. Wan, A. M. Turner, A. Vishwanath, and S. Y. Savrasov, Phys. Rev. B {\bf 83}, 205101 (2011).

\bibitem{Abrikosov1971}A. A. Abrikosov and S. Beneslavskii, Sov. Phys. JETP {\bf 32}, 699 (1971).

\bibitem{Lee2012}F. Wang and D.-H. Lee, Phys. Rev. B {\bf 86}, 094512 (2012).

\bibitem{Young2012}S. M. Young, S. Zaheer, J. C. Y. Teo, C. L. Kane, E. J. Mele, and A. M. Rappe, Phys. Rev. Lett. {\bf 108}, 140405 (2012).

\bibitem{Burkov2011}A. A. Burkov and L. Balents, Phys. Rev. Lett. {\bf 107}, 127205 (2011).


\bibitem{Wang2012}Z. Wang, Y. Sun, X.-Q. Chen, C. Franchini, G. Xu, H. Weng, X. Dai, and Z. Fang, Phys. Rev. B {\bf 85}, 195320 (2012).

\bibitem{Liu2014a}Z. K. Liu, B. Zhou, Y. Zhang, Z. J. Wang, H. M. Weng, D. Prabhakaran, S.-K. Mo, Z. X. Shen, Z. Fang, X. Dai, Z. Hussain, and Y. L. Chen, Science {\bf 343}, 864 (2014).

\bibitem{Wang2013}Z. Wang, H. Weng, Q. Wu, X. Dai, and Z. Fang, Phys. Rev. B {\bf 88}, 125427 (2013).

\bibitem{Liu2014b}Z. K. Liu, J. Jiang, B. Zhou, Z. J. Wang, Y. Zhang, H. M. Weng, D. Prabhakaran, S-K. Mo, H. Peng, P. Dudin, T. Kim, M. Hoesch, Z. Fang, X. Dai, Z. X. Shen, D. L. Feng, Z. Hussain, and Y. L. Chen, Nature Mater. {\bf 13}, 677 (2014).

\bibitem{Marques2017} Y. Marques, A. E. Obispo, L. S. Ricco, M. de Souza, I. A. Shelykh, and A. C. Seridonio, Phys. Rev. B {\bf 96}, 041112(R) (2017).

\bibitem{Jaklevic1966}R. C. Jaklevic and J. Lambe, Phys. Rev. Lett. {\bf 17}, 1139 (1966).

\bibitem{Lambe1968}J. Lambe and R. C. Jaklevic, Phys. Rev. {\bf 165}, 821 (1968).

\bibitem{Zhu2003}J.-X. Zhu and A. V. Balatsky, Phys. Rev. B {\bf 67}, 165326 (2003).

\bibitem{Haug1996}H. Haug and A.P. Jauho, {\it Quantum Kinetics in Transport and Optics of Semiconductors} (Springer, New York, 1996).

\bibitem{Lang1963}I. G. Lang and Yu. A. Firsov, Sov. Phys. JETP {\bf 16}, 1301 (1963).

\bibitem{Mahan2000}G. D. Mahan, {\it Many-Particle Physics} (Kluwer Academic/Plenum Publishers, New York, 2000).

\bibitem{Hubbard1964}J. Hubbard, Proc. R. Soc. London A {\bf 281}, 401 (1964).


(1981).
\bibitem{Borisenko2014}S. Borisenko, Q. Gibson, D. Evtushinsky, V. Zabolotnyy, B. B\"uchner, and R. J. Cava, Phys. Rev. Lett. {\bf 113}, 027603 (2014).

\bibitem{Neupane2014}M. Neupane, S.-Y. Xu, R. Sankar, N. Alidoust, G. Bian, C. Liu, I. Belopolski, T.-R. Chang, H.-T. Jeng, H. Lin, A. Bansil, F. Chou, and M. Z. Hasan, Nat. Commun. {\bf 5}, 3786 (2014).

\bibitem{Hewson1979}A. C. Hewson and D. M. Newns, J. Phys. C: Solid St. Phys. {\bf 12}, 1665 (1979).

\bibitem{ground}We introduce the molecule ground state as the lowest energy level of the DOS spectrum, that is self-consistently obtained in temperature-dependent numerical calculation via occupation number. This counterintuitive statement lies on the fact that for a pair of localized impurities the formation of the lowest energy level is correlations-driven effect, which is being rather sensitive to temperature. Thus, our definition of ground state is different from that used in many-body theory at $T=0$.

\bibitem{Doty2009} M. F. Doty, J. I. Climente, M. Korkusinski, M. Scheibner, A. S.
Bracker, P. Hawrylak, and D. Gammon, Phys. Rev. Lett. {\bf 102},
047401 (2009).

\bibitem{Yakimov2011}A. I. Yakimov, V. A. Timofeev, A. I. Nikiforov, and A. V. Dvurechenskii, JETP Lett. {\bf 94}, 744 (2011).


\end{thebibliography}
\end{document}